\documentclass[preprint,superscriptaddress,showpacs,preprintnumbers,floatfix, amsmath,amssymb]{revtex4}
\usepackage{graphicx}
\usepackage{graphics,amssymb,amsmath,epsfig,color} % % added 
\begin{document}
\title{Time-domain response of atomically thin $\mathrm{MoS_2}$ nanomechanical resonators}
\author{R. van Leeuwen}
\affiliation{Kavli Institute of Nanoscience, Delft University of Technology, Lorentzweg 1, 2628 CJ Delft, The Netherlands}
\author{A. Castellanos-Gomez}
\affiliation{Kavli Institute of Nanoscience, Delft University of Technology, Lorentzweg 1, 2628 CJ Delft, The Netherlands}
\author{G.A. Steele}
\affiliation{Kavli Institute of Nanoscience, Delft University of Technology, Lorentzweg 1, 2628 CJ Delft, The Netherlands}
\author{H.S.J. van der Zant}
\affiliation{Kavli Institute of Nanoscience, Delft University of Technology, Lorentzweg 1, 2628 CJ Delft, The Netherlands}
\author{W.J. Venstra}\email{w.j.venstra@tudelft.nl}
\affiliation{Kavli Institute of Nanoscience, Delft University of Technology, Lorentzweg 1, 2628 CJ Delft, The Netherlands}
\date{\today}
\pagebreak
\begin{abstract} We measure the energy relaxation rate of single- and few-layer molybdenum disulphide ($\mathrm{MoS_2}$) nanomechanical resonators by detecting the resonator ring-down. Recent experiments on these devices show a remarkably low quality (Q)-factor when taking spectrum measurements at room temperature. The origin of the low spectral Q-factor is an open question, and it has been proposed that besides dissipative processes, frequency fluctuations contribute significantly to the resonance line-width. The spectral measurements performed thus far however, do not allow one to distinguish these two processes. Here, we use time-domain measurements to quantify the dissipation. We compare the Q-factor obtained from the ring-down measurements to those obtained from the thermal noise spectrum and from the frequency response of the driven device. In few-layer and single-layer $\mathrm{MoS_2}$ resonators the two are in close agreement, which demonstrates that the spectral line-width in $\mathrm{MoS_2}$ membranes at room temperature is limited by dissipation, and that excess spectral broadening plays a negligible role.
\end{abstract}
\maketitle
\indent\indent Micro- and nanomechanical resonators are of interest for a wide range of applications, including sensors for mass, force and pressure, as well as digital devices such as switches and memories. Recently, 2-dimensional layered materials, like graphene and molybdenum disulfide ($\mathrm{MoS_2}$) have been deployed as materials for nanomechanical resonators~\cite{bunch07,chen09,lee13,gomez13,chen13}. These freely suspended devices, which can be  as thin as a single atom, provide the ultimate limit in mass and surface-to-volume ratio, while maintaining a high resonance frequency due to a very high elastic modulus~\cite{lee08,castellanos12am}.  An important parameter of the resonator is the dissipation~\cite{cleland02,unterreithmeier10,cole11,wilson-rae11,eichler11,gomez13,rieger14}, which is often expressed by the quality (Q)-factor. It is defined as the energy stored in the resonator divided by the energy dissipated per cycle. In experimental work with 2-dimensional crystal resonators reported until now, the Q-factor has been determined from a swept-frequency measurement, by normalizing the full-width-at-half-maximum of the resonance peak to the resonance frequency. These spectral measurements invariably show a low Q-factor when performed at room temperature, and the origin of this low spectral Q-factor is an open question.\\ 
\indent\indent 2-Dimensional resonators are strongly nonlinear and very sensitive to fluctuations. In general not only the displacement fluctuates, but also one or more of the resonator parameters. For example, the mass of the resonator changes due to the adsorption and desorption of ambient molecules~\cite{cleland02,atalaya11}. A nonlinear spring constant converts the displacement fluctuations, that arise from thermal noise, to frequency fluctuations. Furthermore, the resonance frequency may fluctuate as a result of interactions with a noisy system, such as a capacitively coupled gate electrode, or a coupled vibrational mode that executes thermal motion~\cite{westra10,mahboob12,venstra12,castellanos12}. While these parametric fluctuations are in principle not dissipative, they contribute to the broadening of the spectral line-width.\\
\indent\indent Theoretical work has suggested that the large spectral line-width in 2-dimensional resonators could arise from such frequency fluctuations, rather than from energy loss through dissipation~\cite{barnard12,croy12, midtvedt14}. This has also been suggested in recent experimental work, in which the temperature-dependence of the spectral line-width of graphene resonators is investigated\cite{miao14}. However, the spectral techniques used so far in itself do not distinguish dissipation from the proposed spectral broadening mechanisms, and therefore the role of non-dissipative broadening effects remains unclear.\\
\indent\indent In this Letter, we present a method to quantify the dissipation of nanomechanical resonators, by measuring the relaxation time. While ring-down measurements have been used before in high-stress micromechanical strings and beams with ring-down times on the scale of a second~\cite{verbridge08,schmid2011,bui12,leeuwen13}, here we study low-Q devices with the thickness of a few atoms, for which a ring-down occurs within a microsecond. This timescale, combined with the relatively weak signal from such membranes, makes the detection of the ring-down signal challenging. A homodyne mixing technique is used, and the measured response is averaged in the time-domain, which allows time-resolved measurements on micrometer-scale devices with atomic thickness. The Q-factors that correspond to the relaxation time, $\mathrm{Q_R}$, are compared with the spectral Q-factors, $\mathrm{Q_S}$, obtained from driven frequency response measurements and from the thermal noise spectrum. The measurements are performed at room temperature ($\mathrm{300\,K}$) on micrometer-sized drum resonators fabricated from suspended single-layer and few-layer $\mathrm{MoS_2}$ crystals with low residual stress, in which, compared to graphene, a lower dissipation is expected due to a large energy gap in the phonon dispersion~\cite{jiang14}.\\ 
\begin{figure}[h]
\includegraphics[width=85mm]{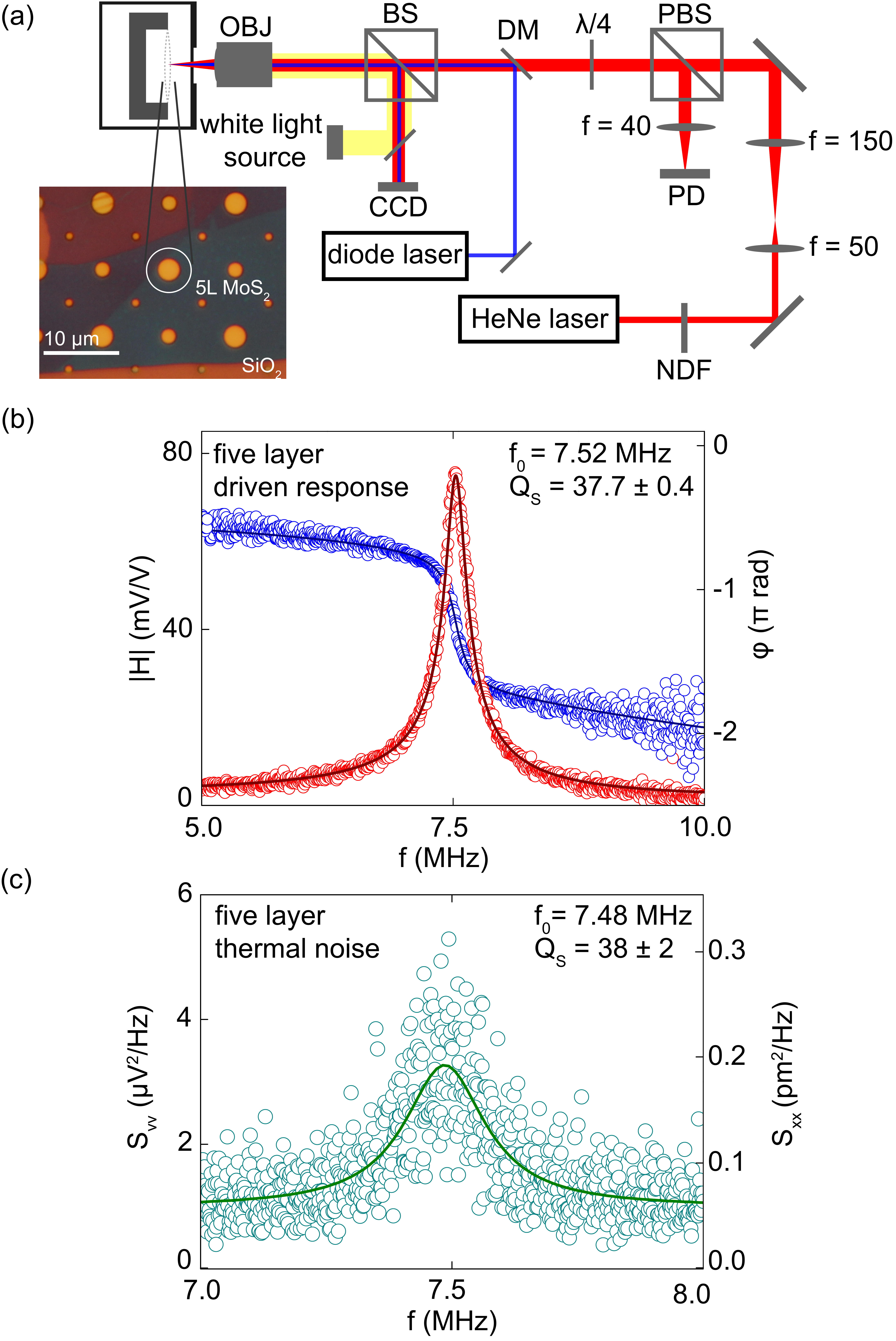}
\caption{(a) Schematic of the optical interferometer.  NDF: neutral density filter; PBS: polarising beam splitter; PD: high-speed photodetector; $\lambda/4$: quarter wave plate; DM: dichroic mirror; BS: beam splitter; CCD: camera; OBJ: objective lense. (inset) Top-view photograph of a 5-layer $\mathrm{MoS_2}$ drum on top of a substrate that was pre-patterned with holes. The circle marks a drum with a diameter of $\mathrm{3\,\mu m}$, which is measured in (b) and (c). (b) Frequency response (magnitude $\mathrm{\vert H\vert}$ and phase $\mathrm{\phi}$) of the weakly driven drum. The solid lines represent a damped-driven harmonic oscillator, with a resonance frequency $\mathrm{f_0 = 7.52\, MHz}$ and a Q-factor of $\mathrm{Q_S= 37.7\pm 0.4}$. (c) Thermal noise spectrum of the drum; solid line: Lorentzian fit with  $\mathrm{f_0 = 7.48\,MHz}$ and $\mathrm{Q_S= 38\pm 2}$.}
\end{figure}
\indent\indent  To fabricate the nanomechanical drum resonators, we start with a boron-doped silicon wafer with a 285 nm thick layer of thermally grown silicon oxide. Circular holes with a diameter ranging from 1 to $\mathrm{5\,\mu m }$ are patterned in the silicon oxide using electron-beam lithography and subsequent dry etching. Thin layers of $\mathrm{MoS_2}$ are obtained by mechanically exfoliating bulk crystals onto a poly(dimethylsiloxane) (PDMS) film. The flakes are then transferred onto the substrate using a recently developed all-dry transfer technique with a micromanipulator and a microscope~\cite{gomez13b}. Figure 1(a), inset, shows a $\mathrm{MoS_2}$ flake transferred onto the pre-patterned substrate.\\ 
\indent\indent The drum motion is probed using the optical interferometer~\cite{bunch07,gomez13} shown schematically in Fig.~1(a). The silicon substrate acts as the fixed mirror, while the suspended nanodrum acts as the moving mirror. A $\mathrm{25\,mW}$ linearly polarized Helium-Neon laser, whose output power is attenuated through a neutral density filter, is expanded ($\mathrm{3\times}$) using two lenses, as to match the aperture of the objective lens ($\mathrm{50\times,\,numerical\,aperture = 0.60,\,working\,distance = 9 \,mm}$) to obtain a diffraction-limited spot. Before hitting the drum, the light passes a polarizing beam splitter (PBS) and a quarter-wave plate. The reflected light passes again through the $\mathrm{\lambda/4}$ plate and is directed by the PBS onto the photodetector. The resonator is driven photothermally using the light from a blue diode laser with an rf-modulated intensity, that is coupled in via the dichroic mirror. During the non-driven measurements, the modulation depth is set to zero. All measurements are conducted at room temperature and, to avoid damping from viscous air, at a pressure of $\mathrm{\sim 10^{-3}\,mbar}$.\\ 
\indent\indent  We start with a measurement of the spectral line-width, by measuring a driven frequency response using a network analyzer. This is the typical measurement from which the Q-factor of nanomechanical resonators is determined. Figure 1(b) shows the amplitude and phase response of the drum encircled in Fig.1(a), with a diameter of $\mathrm{3\,\mu m}$ and a thickness of 5 layers ($\sim$ 3.2 nm). The response of the weakly driven fundamental mode fits well to a damped-driven harmonic oscillator, with a resonance frequency $\mathrm{f_{0} = 7.52\,MHz}$ and a Q-factor $\mathrm{Q_S=37.7\pm0.4}$.  This spectral Q-factor is typical for resonators made from 2-D materials at room temperature~\cite{eichler11,barton11,gomez13}, and is in sharp contrast to the Q-factors observed for similar devices at low temperatures which can exceed $\mathrm{10^5}$~\cite{huttel09, weber14, singh14}. Recent work shows that the line-width in resonators from suspended carbon nanotubes and graphene could depend on the amplitude of the motion~\cite{lifshitz08,eichler11}. To verify whether the driving force gives rise to spectral broadening, we measure the motion of the drum without driving force. Figure 1(c) shows the thermal noise spectrum. From a Lorentzian fit, we find $\mathrm{f_0 = 7.48\,MHz}$ and $\mathrm{Q_S= 38\pm 2}$, which is in close agreement with the driven measurement. Furthermore, no significant change in the spectral Q-factor was found when the drive strength was varied~\cite{supmat}. From this we conclude that driving-force dependent spectral broadening plays no significant role in our device.\\ 
\begin{figure}[h]
\includegraphics[width=80mm]{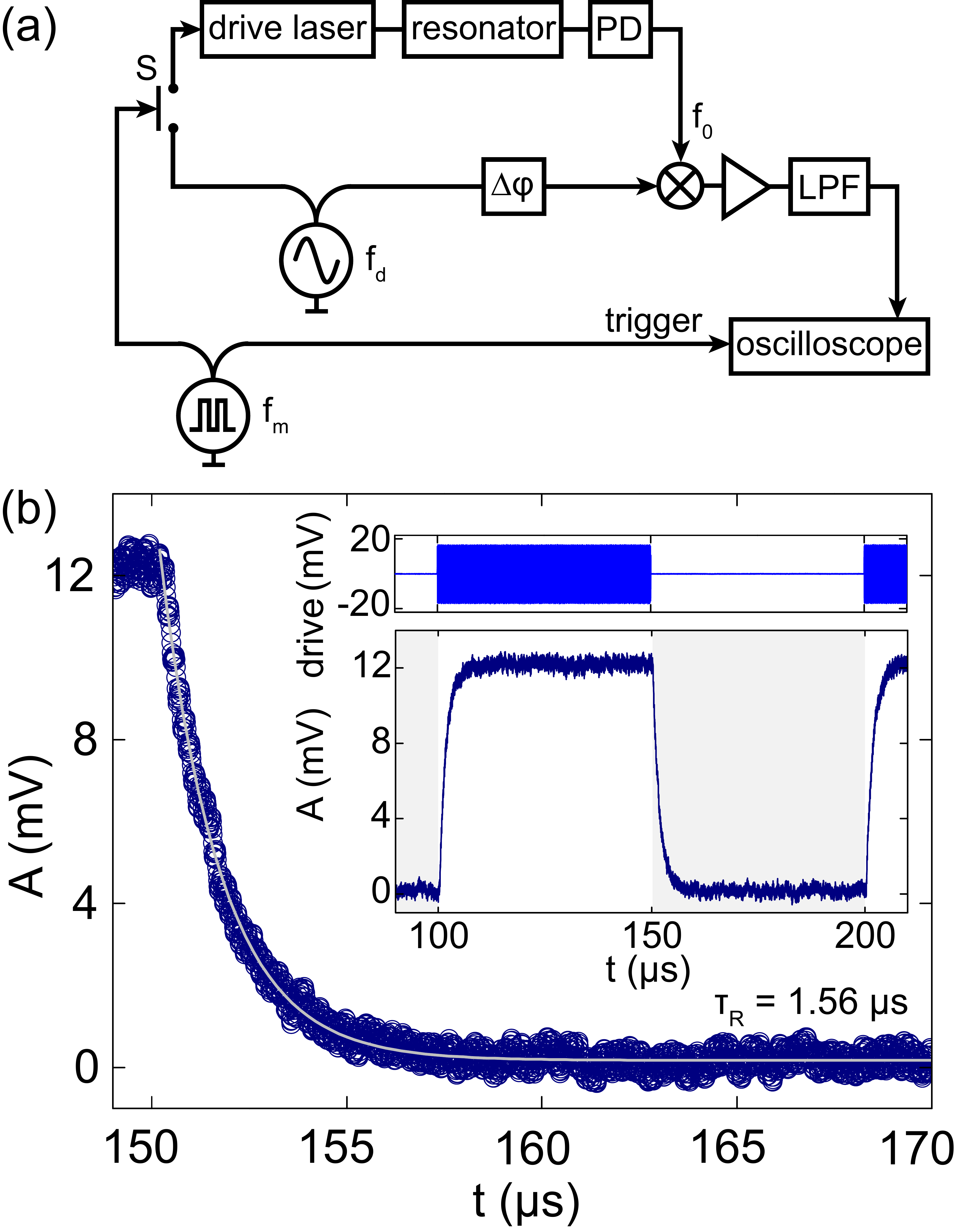}
\caption{(a) Schematic of the ring-down measurement. The drive signal at $\mathrm{f_{d}}$ is a applied to the sample via rf-switch S, and used as the reference oscillator in the detector. During the ring down the drum oscillates at $\mathrm{f_0}$. The switch is controlled using a pulse with a $\mathrm{50\,\%}$ duty cycle at $\mathrm{f_m=10\,kHz}$. (b) Time-domain measurements on the five-layer $\mathrm{MoS_2}$ drum. (inset) Driving signal and drum response, measured with $\mathrm{f_d=f_0= 7.52 MHz}$. (main panel) Averaged amplitude trace and fitted exponential function with $\mathrm{\tau_R= 1.56\,\mu}$s, which corresponds to a Q-factor $\mathrm{Q_R=37.8\pm 0.2}$.} 
\end{figure} 
\indent\indent To measure the mechanical ring-down, we introduce the measurement scheme depicted in Fig. 2(a). The drum is driven  at $\mathrm{f_d}$ using the laser diode, and its motion is probed by the photodiode. By opening the rf-switch the driving signal is interrupted, which causes the resonator to relax to equilibrium. The quadrature corresponding to the drum amplitude, which is selected using the phase shifter, is recorded on an oscilloscope. The time-domain signal is averaged over multiple ring-down events (typically 1000), by triggering the scope in synchrony with the rf-switch. In the present experiments the trigger frequency is $\mathrm{10\,kHz}$, and the duty-cycle of the trigger pulse equals $50\%$, enabling the detection of the resonator ring-down and the ring-up transients. The relaxation times are obtained by fitting an exponential function to the averaged amplitude data, with  $\mathrm{y = y_0+A_1 e^{-((t-t_0)/\tau_R)}}$, and the ring-down Q-factor is then calculated as $\mathrm{Q_R=\pi\tau_Rf_0}$. Since the ring-down measurement is averaged in the time domain, it is insensitive to frequency fluctuations~\cite{cleland02,quadrature,schneider14}. This makes it possible to distinguish line-width broadening by dissipative processes from those occurring from parametric noise.\\ 
\indent\indent Figure 2(b) shows a time trace measured with $\mathrm{f_d = f_ 0 = 7.52 MHz}$, for the five-layer thick drum resonator discussed in Fig. 1(b,c). The top panel of the inset shows the driving signal, the bottom panel the ring-up and ring-down signal after $\mathrm{1000\times}$ averaging. When the resonator is driven off-resonance, $\mathrm{f_d \neq f_0}$, and then the driving force is switched off, the ring-down occurs at the resonance frequency. This causes the detected time-domain signal to oscillate at the detuning frequency $\mathrm{f_d - f_0}$~\cite{supmat}. Figure 2(b), main panel, shows the ring-down data for $\mathrm{f_d = f_0}$. From the exponential fit we extract $\mathrm{1.56\,\mu s}$, which corresponds to a ring-down Q-factor $\mathrm{Q_R = 37.8\pm 0.2}$. This value is in close agreement with the spectral Q-factors measured earlier. These measurements demonstrate that for five-layer mechanical resonators at room temperature the spectral line-width is dominated by dissipative processes, and that dispersive line-width-broadening processes play a minor role.\\
\begin{figure}[h]
\includegraphics[width=80mm]{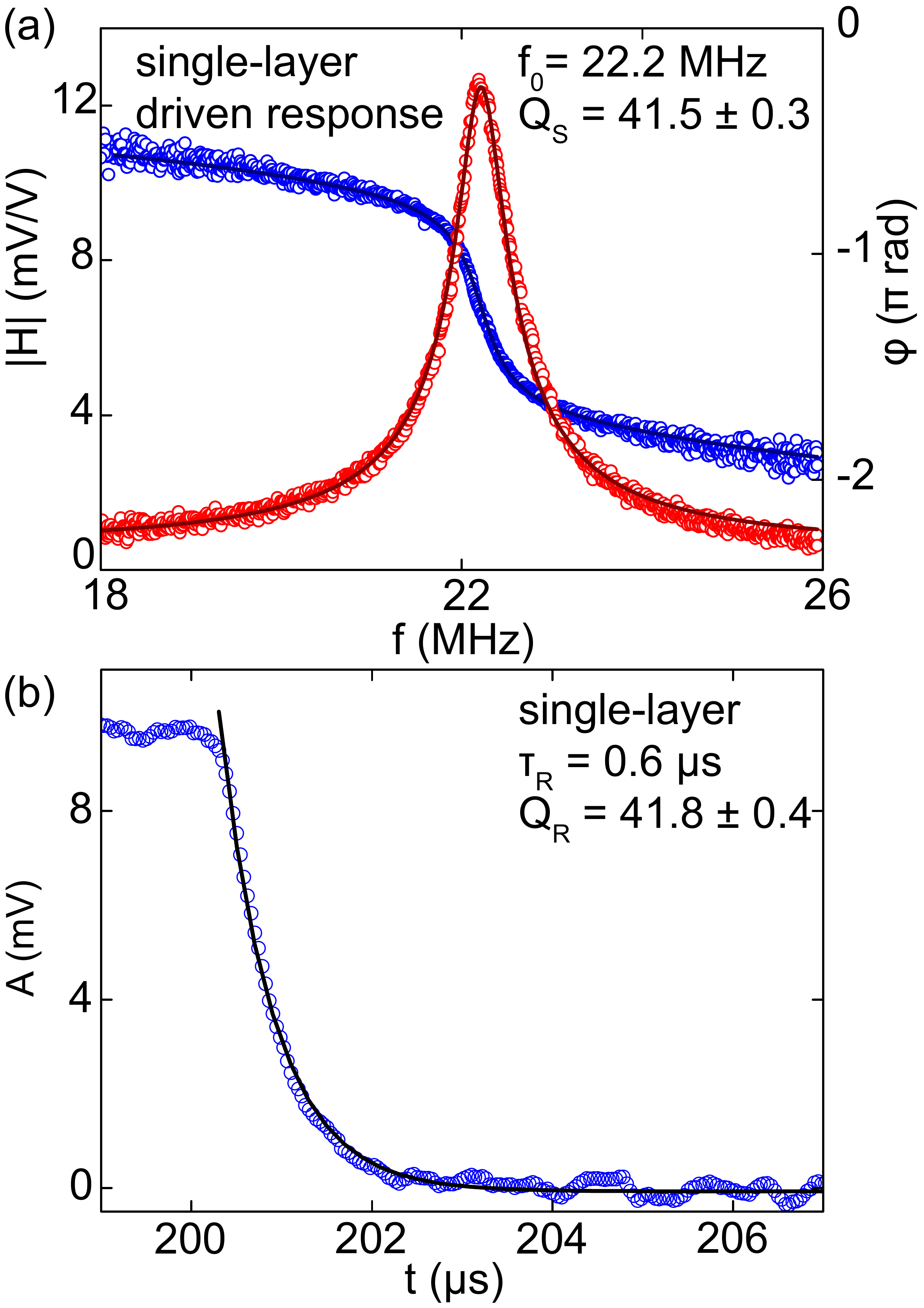}
\caption{(a) Frequency response (magnitude $\mathrm{\vert H\vert}$ and phase $\mathrm{\phi}$) of a single-layer $\mathrm{MoS_2}$ resonator with a diameter of $\mathrm{2\,\mu m}$, and a harmonic oscillator fit with $\mathrm{f_0 = 22.2\,MHz}$ and the $\mathrm{Q_S = 41.5\pm 0.3 }$. (b) Ring-down measurement of the same device, and exponential fit with $\mathrm{\tau_R=0.6\,\mu s}$, which corresponds to $\mathrm{Q_R = 41.8\pm0.4} $.}
\end{figure}
\indent\indent While five-layer $\mathrm{MoS_2}$ resonators vibrating in the linear regime mechanically behave almost like a plate, thinner devices are in the membrane regime~\cite{gomez13}. In a membrane, the bending rigidity becomes negligible and the restoring force arises from the initial tension and the displacement-induced tension. In such devices, a fluctuating tension could give rise to significant frequency fluctuations, leading to a low $\mathrm{Q_S}$~\cite{barnard12,venstra12,eriksson13}. To investigate whether spectral broadening can be observed in the membrane regime, we measure the spectral and ring-down Q-factors of a single-layer $\mathrm{MoS_2}$ membrane, with a diameter of  $\mathrm{2\,\mu m}$.  Figure 3(a) shows the amplitude and phase response of the weakly driven device, and the fitted harmonic oscillator. We extract $\mathrm{f_0 = 22.2\,MHz}$ and $\mathrm{Q_S = 41.5\pm 0.3}$. Figure 3(b) shows a ring-down measurement of the same device, with $\mathrm{f_d = f_0 = 22.2\,MHz}$. From the exponential fit, the ring-down time is  $\mathrm{\tau_R = 0.6 \mu s}$, which corresponds to $\mathrm{Q_R = 41.8\pm0.4}$. This is in agreement with the spectral Q-factor, from which we conclude that also for the suspended single-layer resonator, which is in the membrane limit, the spectral line-width is limited by dissipative processes.\\
\indent\indent These experiments show that for nanomechanical resonators fabricated from 2-dimensional materials, the Q-factors obtained from spectral measurements at room temperature are in good agreement with those determined from ring-down experiments. This observation applies to few- and single-layer resonators, that vibrate in the plate and in the membrane regimes. This indicates that parametric spectral broadening is not the main cause of the experimentally observed low Q-factors in such devices. Possible candidates for the low Q-factor should be sought in dissipative mechanisms, such as surface effects and clamping losses. Given the high Q-factors observed at low temperatures, it will be interesting to investigate how the dissipation develops as a function of the temperature. To this end, spectral line-width measurements~\cite{miao14} could be complemented by measurements of the relaxation time, using the presented measurement technique. On the other hand, more insight in the effects of parametric spectral broadening could be obtained by applying controllable parametric noise to the resonance mode, for example by driving a gate electrode or a different vibration mode of the same resonator with a stochastic signal. Using the detection scheme presented here to measure the relaxation time of the resonator, which may be on the timescale of a microsecond, the dispersive and dissipative mechanisms that contribute to the width of the resonance peak can then be discriminated.\\
\indent\indent The authors acknowledge financial support from NanoNextNL, a micro and nanotechnology consortium of theNetherlands and 130 partners, the FP7-Marie Curie Project PIEF-GA-2011-300802 (STRENGTHNANO), and the European Union's Seventh Framework Programme (FP7) under Grant Agreement $\mathrm{n{\circ}~318287}$, project LANDAUER.\\
%\bibliographystyle{apsrev}
%\bibliography{ringdown}

%%%%%%%%%% Merge with supplemental materials %%%%%%%%%%
\pagebreak
\widetext
\begin{center}
\textbf{\large Supplemental Material to:\\ Time-domain response of atomically thin $\mathrm{MoS_2}$ nanomechanical resonators}\\
R. van Leeuwen, A. Castellanos-Gomez G.A. Steele,\\ H.S.J. van der Zant, W.J. Venstra\\
\emph{Kavli Institute of Nanoscience, Delft University of Technology,\\ Lorentzweg 1, 2628 CJ Delft, The Netherlands}\\w.j.venstra@tudelft.nl
\end{center}
%%%%%%%%%% Merge with supplemental materials %%%%%%%%%%
%%%%%%%%%% Prefix a "S" to all equations, figures, tables and reset the counter %%%%%%%%%%
\setcounter{equation}{0}
\setcounter{figure}{0}
\setcounter{table}{0}
\setcounter{page}{1}
\makeatletter
\renewcommand{\theequation}{S\arabic{equation}}
\renewcommand{\thefigure}{S\arabic{figure}}
\renewcommand{\bibnumfmt}[1]{[S#1]}
\renewcommand{\citenumfont}[1]{S#1}
%%%%%%%%%% Prefix a "S" to all equations, figures, tables and reset the counter %%%%%%%%%%

\indent\indent\begin{figure}[h]
\includegraphics[width=170mm]{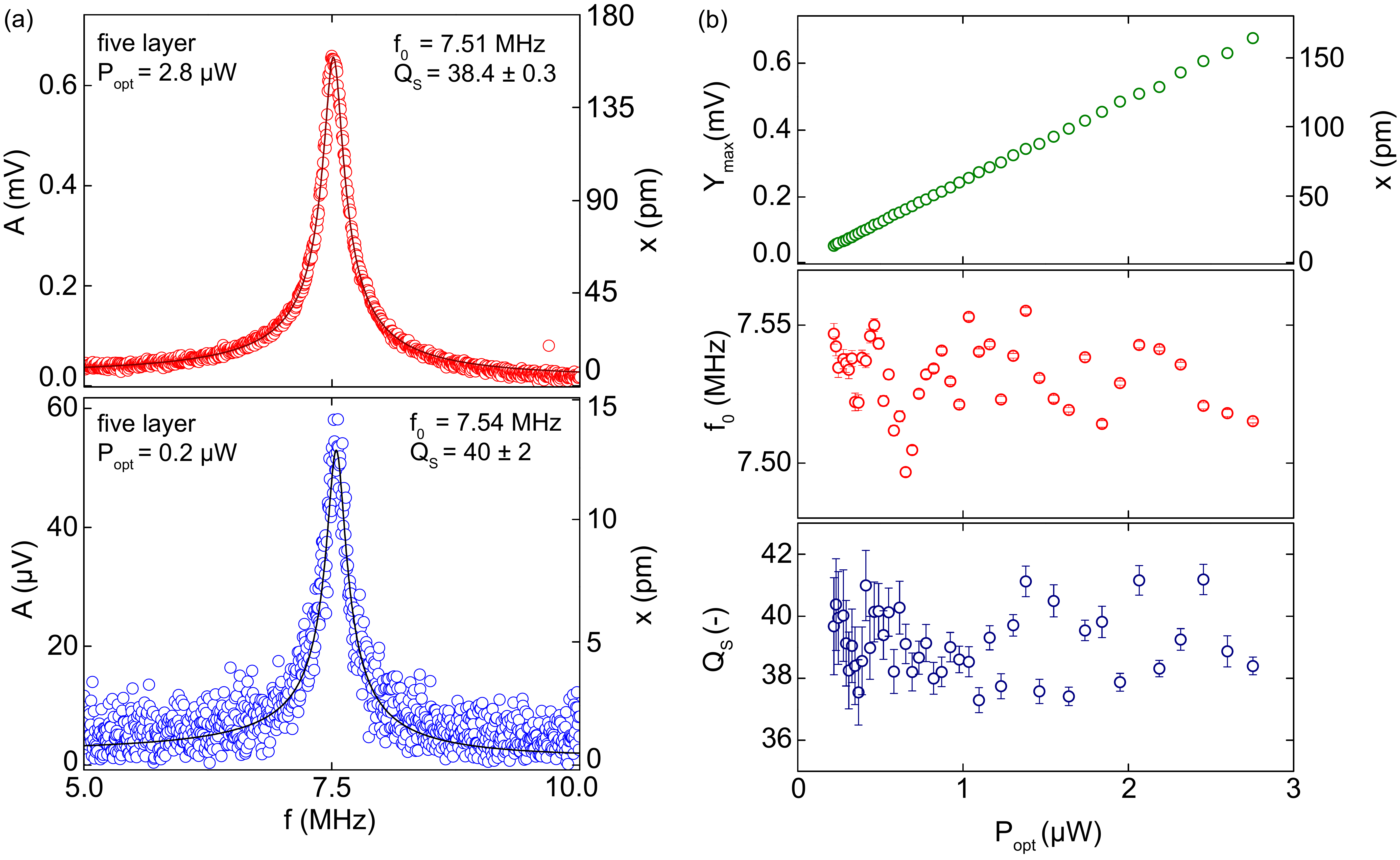}
\caption{Dependence of  $\mathrm{Q_S}$ on the amplitude of the motion for the five-layer resonator with a diameter of $\mathrm{3\,\mu m}$ of Fig. 1, main text. (a) Frequency response (amplitude shown only) at weak (bottom panel) and strong (top panel) driving. The solid line is a harmonic oscillator function fit to the magnitude and phase data. The incident ac optical power, $\mathrm{P_{opt}}$,  is indicated in the insets, and the secondary y-axis indicates the resulting mechanical displacement. The solid lines represent damped-driven harmonic oscillator fits, from which $\mathrm{f_0}$ and $\mathrm{Q_S}$ are obtained. (b) Top panel: detected photodiode signal and corresponding drum displacement when varying the optical power, demonstrating a linear relation between the driving force and the amplitude of the resulting motion. Center panel: resonance frequency as a function of the driving power. Bottom panel: spectral Q-factor, $Q_{S}$, as a function of the driving power. Compared to the spectral measurements presented in Figures 1 and 2 in the main text no significant change in $Q_{S}$ is observed, which indicates that over this range the dephasing has a negligible effect on the spectral Q-factor}
\end{figure}
\newpage 

\begin{figure}[h]
\includegraphics[width=80mm]{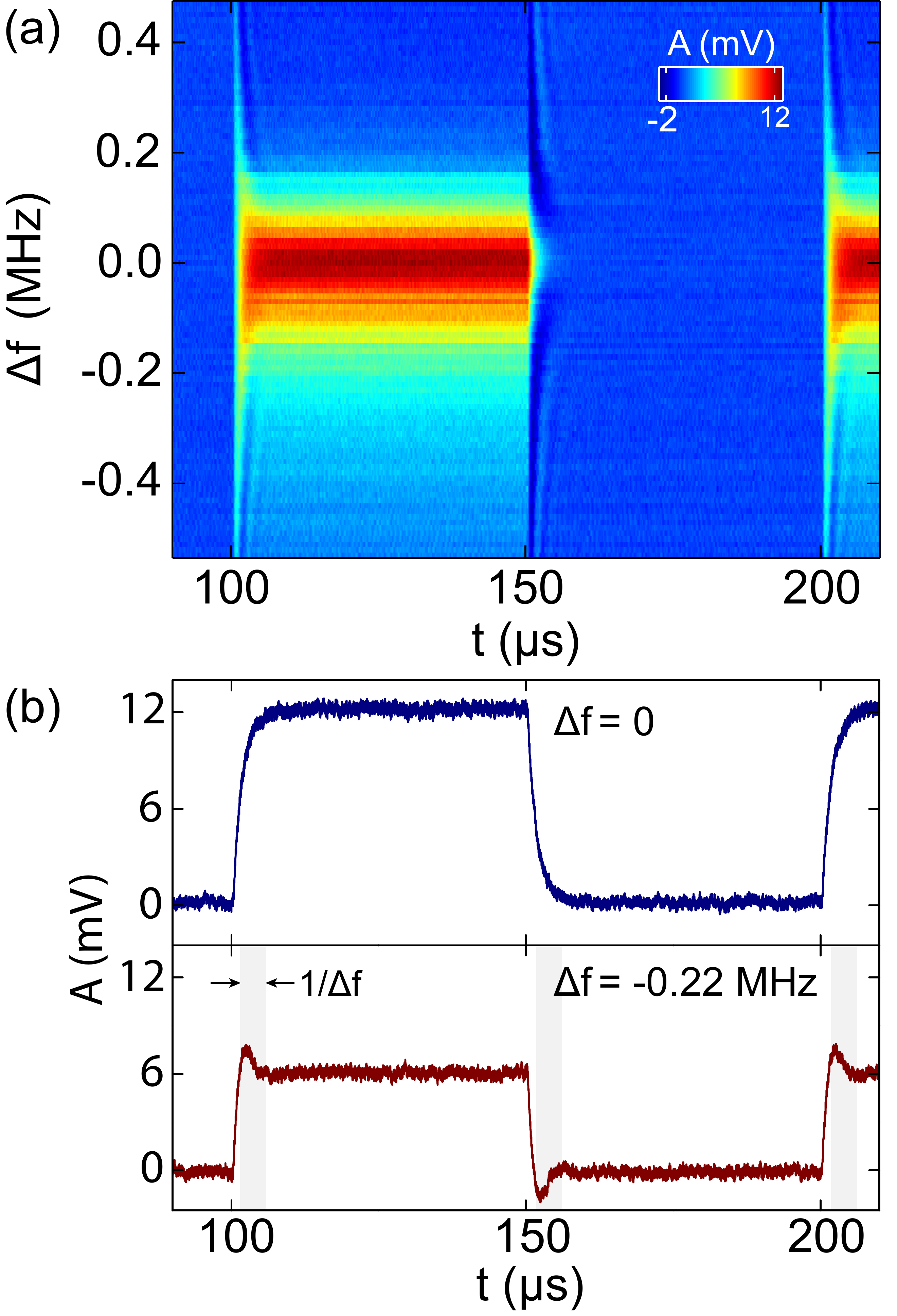}
\caption{Ring-down measurements with a detuned reference oscillator. (a) Surface plot showing the $\mathrm{1000\times}$-averaged time-domain traces (horizontal, fast axis) while varying the detuning $\mathrm{\Delta f= f_d-f_0}$ (vertical, slow axis). During ringdown, the mechanical resonator oscillates at its natural frequency, while the reference for the homodyne detection is still given by the driving frequency. For ring-down traces measured with the driving frequencies detuned from the resonator natural frequency, this results in a oscillation of the homodyne signal at the difference frequency $\mathrm{\Delta f}$. (b) Cross-sections taken on-resonance (upper panel,  $\mathrm{\Delta f = 0}$), and with detuning (lower panel, $\mathrm{\Delta f = -0.22\,MHz}$). The transient response oscillates with a period of $\mathrm{1/\Delta f = 4.5\, \mu s}$, as visualized by the grey bars.} 
\end{figure}

\end{document}